\documentclass[5p,times, twocolumn]{elsarticle}

\usepackage{amssymb}
\usepackage{amsmath}
\usepackage{verbatim}
\usepackage{mathrsfs}
\usepackage{amsfonts}
\usepackage{latexsym}
\usepackage{epsfig}
\usepackage{color}
\usepackage{cancel}
\usepackage{graphicx,subfigure}
\usepackage{units}
\journal{Physics Letters B}

\begin{document}
\begin{frontmatter}
\title{Distinguishing Dirac and Majorana neutrinos with astrophysical fluxes}
\author[rvt]{J. Barranco}
\author[rvt]{D. Delepine}
\author[rvt]{M. Napsuciale}
\author[rvt]{A. Yebra}
\address[rvt]{Departamento de F\'isica, Divisi\'on de Ciencias e Ingenier\'ias, Campus Le\'on,
Universidad de Guanajuato, Le\'on 37150, M\'exico}

\begin{abstract}
Massive neutrinos can have helicity $s_{\parallel}\neq -1$. Neutrino helicity changes when the neutrino interacts with an 
external magnetic field and it is possible that
the left-handed neutrinos born inside the Sun or a supernova 
could leave their sources with a different helicity. 
Since Dirac and Majorana neutrinos have different cross sections in the scattering on electrons for different neutrino helicities, 
a change in the final neutrino 
helicity may generate a different number of events and spectra in terrestrial detectors when astrophysical neutrinos 
have travelled regions with strong magnetic fields.
In this work, we show that looking for these effects in solar neutrinos, it could be possible to set  bounds in the neutrino 
properties such as the neutrino magnetic moment. Furthermore, for neutrinos coming from a supernova, we show that 
even in the case of an extremely small neutrino magnetic moment, $\mu_\nu \sim 10^{-19}\mu_B$, there will be 
measurable differences in both the number of events and in the spectra of Majorana and Dirac neutrinos.  
\end{abstract}
\end{frontmatter}
\section{Introduction}
A fundamental challenge faced by the particle physics community is to determine the Majorana or Dirac nature of the 
neutrino. In order to assess this question, experimentalists are exploring different reactions where the Majorana nature 
may be manifested (for a review see e.g. \cite{Zralek:1997sa}).  Well known facts concerning this problem are: 
\begin{enumerate}
\item a  Majorana particle is identical to its own antiparticle and it leads to reactions where the lepton number is not conserved. 
The prototypical example of such processes is the neutrinoless double beta decay 
\cite{Schechter:1981bd,Deshpande:1984sm,DellOro:2016tmg,Vergados:2016hso} and,
\item massive neutrinos can have helicity $s_{\parallel}\neq-1$ and there are helicity-driven effects yielding a sizeable difference in 
the scattering cross sections for Majorana and Dirac  neutrinos on electrons \cite{Singh:2006ad,Kayser:1981nw,Barranco:2014cda}.
\end{enumerate}
Most of the experimental effort is focused in the search for neutrinoless double beta decay. 
Multiple experiments have been constructed for this purpose like the Heidelberg-Moscow experiment \cite{KlapdorKleingrothaus:2000sn}, 
IGEX \cite{Aalseth:2002rf}, EXO \cite{Auger:2012ar}, GERDA \cite{Agostini:2017iyd}, Kamland-ZEN \cite{Gando:2012zm} or 
CUORE \cite{Arnaboldi:2002du}, among others.

The purpose of this work is to show that the second possibility may lead to observable differences between the Majorana 
and Dirac neutrinos. Previous works on this topic have been concentrated in a full conversion of left handed 
Majorana neutrino into right handed Majorana anti-neutrino \cite{Barbieri:1991ed,Semikoz:1996up,Pastor:1997pb}. The 
non observation of electron anti-neutrinos in solar detectors have set strong limits on the neutrino magnetic moment 
\cite{Miranda:2003yh}. Here, we will show that it is not necessary to have a full neutrino-antineutrino conversion in order to obtain a 
positive signal of this effect, but only a change in the vector polarization will lead to measurable differences. 

We organized the paper as follows: In section \ref{cross_section}, we present the Dirac and Majorana neutrino-electron 
elastic scattering cross sections as  functions of the polarization vector of the incident neutrino. Next, in 
section \ref{change} we study the change of neutrino helicity in the presence of a magnetic field. In 
section \ref{sunmagnetic},  assuming a model for the magnetic field of the Sun \cite{Miranda:2000bi}, we study this effect 
for solar neutrinos and translate results of the direct measurement of the $^7$Be solar neutrino signal rate performed 
by Borexino \cite{Arpesella:2008mt} into a constraint for the neutrino magnetic moment.  
Finally, in section \ref{supernova}, 
we discuss the case of supernova neutrinos where, due to the strong magnetic fields generated in a supernova, in some cases 
as strong as $B\sim 10^{15}$ Gauss \cite{Obergaulinger2014,Obergaulinger:2011ic}, even an extremely small neutrino 
magnetic moment, as small as  $\mu_\nu \sim 10^{-19}\mu_B$, can make a significant change in the polarization 
vector, leading to measurable differences between Majorana and Dirac neutrinos in both the number of events and 
the energy spectrum. 
\section{The $\nu$-e scattering cross section including neutrino polarization }\label{cross_section}
Possible differences in the $\nu$-e scattering cross section between Dirac and Majorana neutrinos have been 
previously considered concluding that such differences are proportional to the neutrino masses  
\cite{Dass:1984qc,Garavaglia:1983wh}. Because of the smallness of the neutrino masses, in practice, no measurable 
difference in the Majorana or Dirac $\nu$-e scattering cross section seems to be possible.
Nevertheless, this conclusion applies to the considered unpolarized cross section and, according to 
\cite{Kayser:1981nw,Barranco:2014cda}, when the neutrino polarization is taken into account, 
a clear difference between Majorana and Dirac neutrinos in neutrino-electron scattering appears. Defining the incident 
neutrino polarization vector in the neutrino rest frame as $s_\nu=\left(0,s_\perp,0,s_\parallel\right)$, we can calculate 
the differential cross sections for each case, Dirac and Majorana, in terms of the helicity $s_\parallel$. In 
\cite{Kayser:1981nw,Barranco:2014cda} the cross sections were computed in the center of mass frame. 
Here we present results in the laboratory frame, which is more suitable for the purposes of this paper.  
The differential cross section for $\nu$-$e$ elastic scattering process is given by  

\begin{align}\nonumber
&\frac{d^2\sigma^D}{dE^i_\nu dT}=\frac{m_eG^2_F}{4\pi }\bigg\{(g_A+g_V)^2\frac{E^i_{\nu}(E^{i}_\nu-s_\parallel P)}{P^2}\nonumber\\
&+\left((g_A+g_V)^2\frac{(E^i_\nu-T)^2}{P^2}+(g_A^2-g_V^2)\frac{m_eT}{P^2}\right)\left(1-s_\parallel \frac{E^i_\nu}{P}\right)\nonumber\\
&+\left(\frac{s_\parallel}{P}(g_A-g_V)^2(E^i_\nu-T)\left(1+\frac{T}{m_e}\right)+(g_A^2-g_V^2)\left(1-s_\parallel\frac{T}{P}\right)\right)\,,\nonumber\\
\times&\left(\frac{m_\nu}{P}\right)^2\bigg\}\,,\label{CrossDirac}
\end{align}
in the Dirac case, while if the neutrino is a Majorana particle we have
\begin{align}
\nonumber
&\frac{d^2\sigma^M}{dE^i_\nu dT}=\frac{m_e G^2_F}{2\pi P^2}\bigg\{2\left(2g^2_A-g^2_V\right)m^2_\nu\\ 
& -\frac{4E^i_\nu}{P}g_A g_V s_\parallel T \left(E^i_\nu +\frac{m^2_\nu}{m_e}\right)\left(1-\frac{T}{2E^i_\nu}\right)+\left(g^2_A-g^2_V\right)m_e T\nonumber \\ 
&+\left(g^2_A+g^2_V\right)\left[2{E^i_\nu}^2\left(1-\frac{T}{E^i_\nu}\right)+T^2\left(1+\frac{m^2_\nu}{m_e T}\right)\right]\bigg\},\label{CrossMajorana}
\end{align}
In last two equations,  $P=\left|\vec{P}^i_{\nu}\right|$ is the momentum of the incident neutrino, whereas $T$ represents 
the electron recoil energy. From Eqs. \eqref{CrossDirac} and \eqref{CrossMajorana}, with $m_\nu=0$ and $s_\parallel=-1$, 
the usual differential cross section for $\nu$-$e$ scattering is recovered.
\begin{figure}
\begin{center}
\includegraphics[width=.45\textwidth]{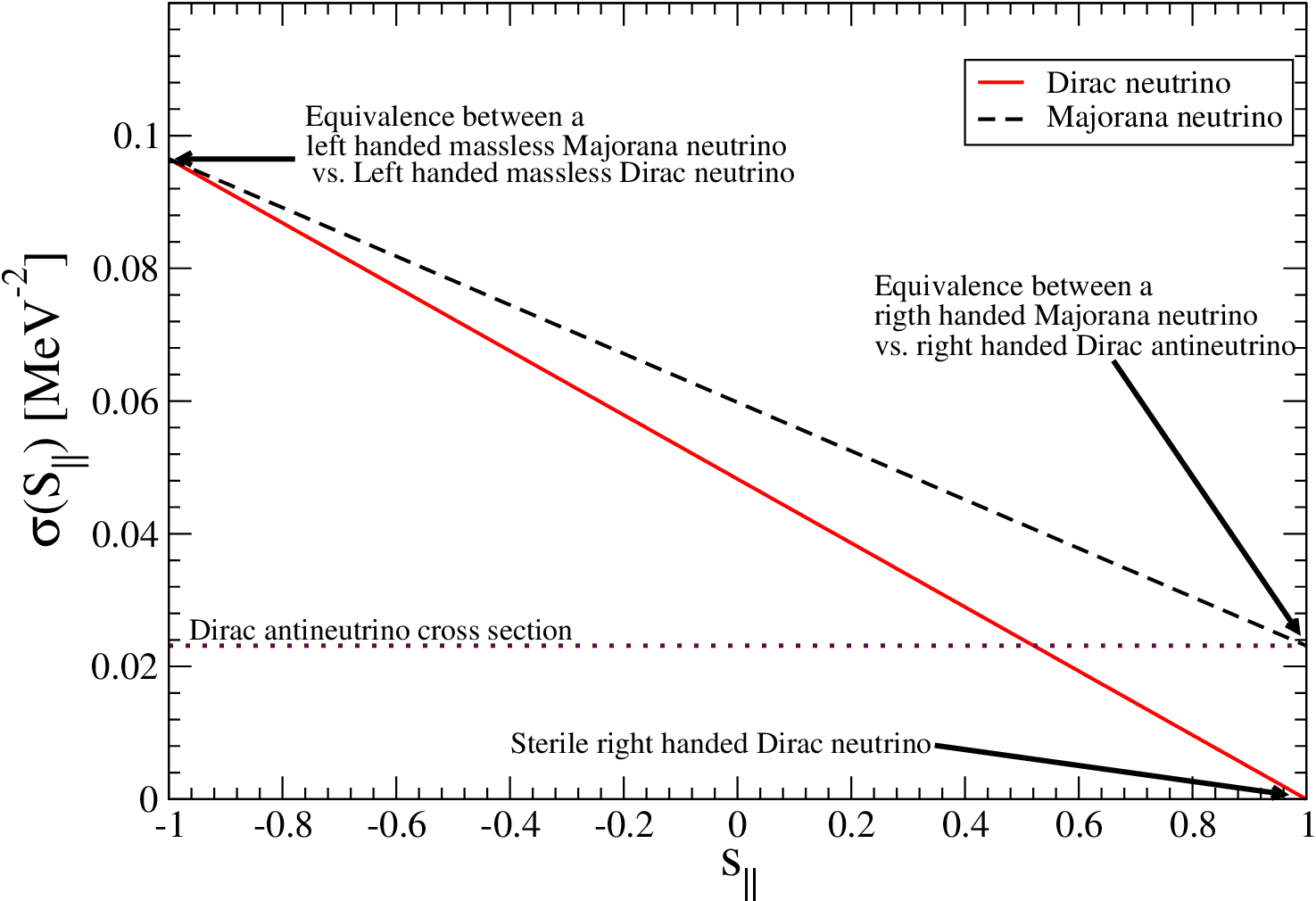}
\caption{Integrated cross section for the Beryllium line as a function of the 
neutrino spin polarization. }\label{Fig1}
\end{center}
\end{figure}
Presently, cosmological data sets a limit on the sum of the neutrino masses $\sum m_\nu < 0.183$ eV 
\cite{Giusarma:2016phn,Ade:2015xua}. Furthermore, 
terrestrial experiments designed to measure the effect of neutrino masses on the tritium $\beta$-decay spectrum near 
its endpoint have set an upper bound on the electron neutrino mass of $m_{\nu_e} < 2.3$ eV at $95\%$ C.L. 
\cite{Kraus:2004zw,Lobashev:2001uu}. Thus, neutrino masses are really small, and the only variable able to produce 
a difference between a Dirac neutrino and a Majorana neutrino is $s_{||}$. Notice that the leading effects for  
$s_{||}\neq -1$ in Eqs. (\ref{CrossDirac}, \ref{CrossMajorana}) are proportional to the electron mass.

In order to illustrate the differences between Dirac and Majorana cross sections driven by $s_{||}$, we calculate the cross 
section for a specific polarization
\begin{equation}
\sigma^{M,D}(s_{||})\equiv \int_{T_{min}}^{T_{max}} dT \int_0^{\infty} dE_\nu \lambda(E_\nu)  \frac{d^2\sigma^{M,D}}{dE_\nu dT}(E_\nu,T,s_{||})\,,
\end{equation}
where $\lambda(E_\nu)$ is the neutrino spectrum, which depends on the neutrino source under consideration. For 
definitiveness, we will use the  $^7$Be line of the solar neutrino spectrum. In this case, the spectrum will 
be a Dirac delta centred in $E_\nu=0.862$ MeV. For the recoil energy $T$, we will assume a detector with the features 
of Borexino, i.e. we consider $T \in [250, 750]$ keV. 

In order to exhibit the size of the helicity-driven effects, in Fig. \ref{Fig1} we plot the integrated cross sections for 
Dirac and Majorana neutrinos taking  
$m_{\nu}=1$ eV in the numeric calculations. We can see three important features from this figure: 
\begin{enumerate}
\item a left handed Dirac neutrino has the same cross section as a left handed Majorana neutrino,
\item a right handed Majorana neutrino has the same cross section as a left handed Dirac antineutrino, and
\item a right handed Dirac neutrino has zero cross section, i.e. it is a sterile neutrino. 
\end{enumerate}

\begin{figure}
\includegraphics[width=0.45\textwidth]{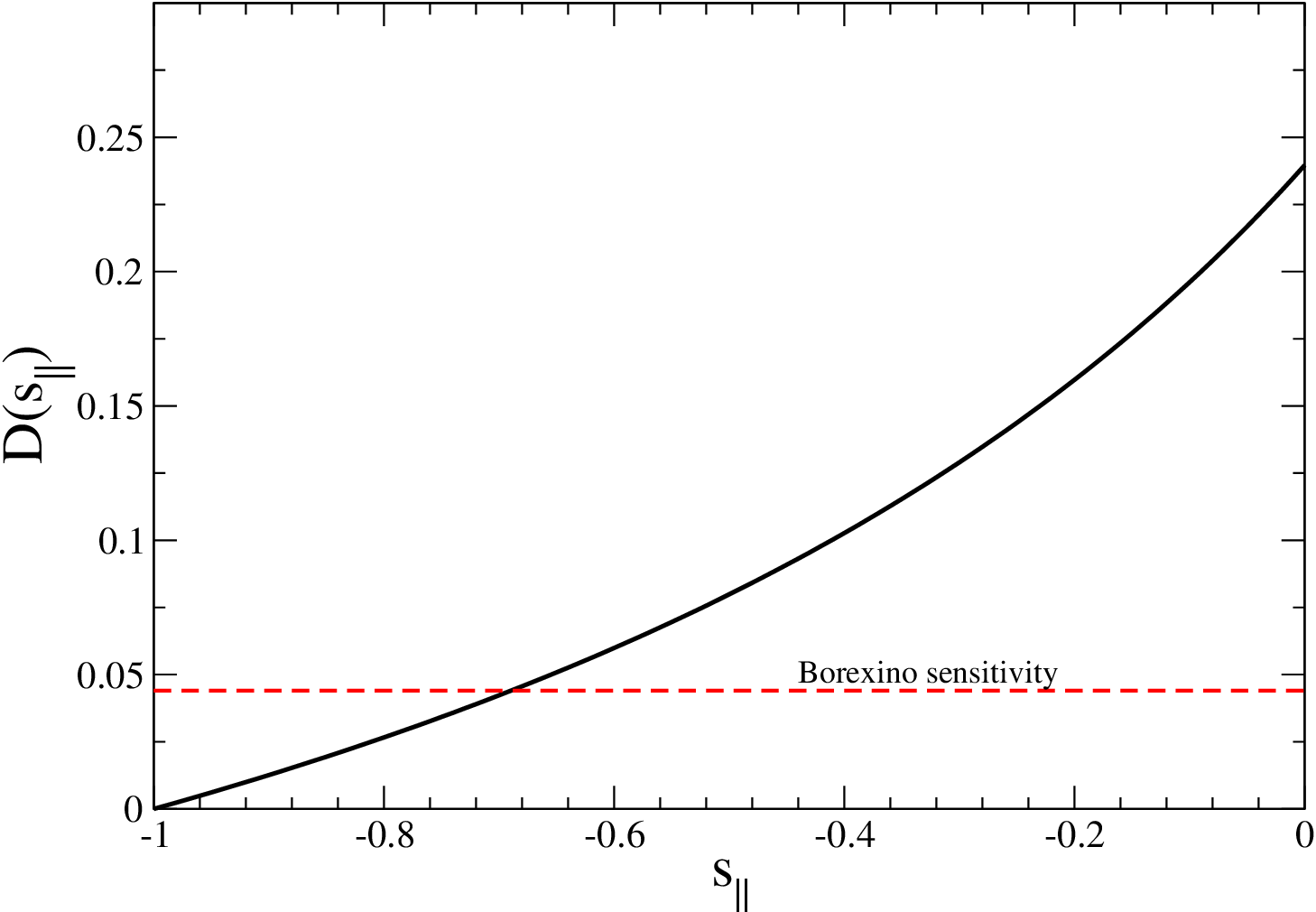}
\caption{The function $D(S_{||})$ for  the $^7$Be line of the solar spectrum within the Borexino integration intervals.}\label{differenceBorexino}
\end{figure}

Besides these three limits, there are differences in the Majorana and Dirac cross sections for -1 $< s_{\parallel} <$ 1. A way 
to quantify this difference is given by means of the function
\begin{equation}
D\left(s_\parallel\right)\equiv\frac{|\sigma^M(s_\parallel)-\sigma^D(s_\parallel)|}{\sigma^D(s_\parallel)}\,,
\label{difference}
\end{equation}
which only depends on the  helicity $s_\parallel$.  We remark that neutrino-mass-driven effects cancel in the difference 
and this function reflects the helicity-driven effects properly. The function $D(s_{||})$ with $\sigma^{M,D}$  integrated 
assuming $\lambda(E_\nu)=\delta(E_\nu-0.862 \textrm{MeV})$ 
and $T \in [0.250,0.750]$ MeV is shown in Fig. \ref{differenceBorexino}. From this plot, we conclude that there is a sizable difference 
in the cross section for the scattering of Majorana and Dirac neutrinos with $s_{||}\neq -1$ off electrons.

Neutrinos are born as left handed-particles. Indeed, for instance, in the decay of a pseudo-scalar meson 
$P^+ \to \ell^+ + \nu$, the neutrino helicity $s_{||}$ can be computed as $s_{||}=(E-W)|\vec k|/(W E-|\vec k|^2)$
with $W$ and $E$ the energies of the charged lepton $\ell$ and the neutrino, respectively \cite{Barenboim:1996cu}. 
Due to the smallness of the neutrino mass, effectively the initial vector polarization can be written as 
$s_\nu=\left(0,s_\perp=0,0,s_\parallel=-1\right)$ and, as we have shown before, Dirac or Majorana $\nu$-$e$ 
cross sections for small neutrino masses and left handed Dirac or Majorana neutrinos  are identical.
The challenge we face now is to find physical processes able to change the neutrino helicity $s_\parallel$. 

\section{Change of neutrino polarization due to a magnetic field}\label{change}
Fortunately, nature offers a way in which the neutrino helicity may be modified. Indeed, any neutral particle with a 
magnetic moment can change its longitudinal part of the polarization vector in the presence of an external magnetic field. 
The change in the helicity is given by the Bargmann-Michel-Telegdi equation \cite{Bargmann:1959gz,Semikoz:1992yw}  
\begin{equation}
\frac{ds_{||}}{dr}=-2\mu_\nu B_\perp s_{||}\,,\label{spinequation}
\end{equation}
where $B_\perp$ is the component of the external magnetic field perpendicular to the propagation of the neutrino 
and $\mu_\nu$ is the effective neutrino magnetic moment.
Neutrinos can have a magnetic moment if they are massive. The effective neutrino magnetic moment is different 
for Majorana and Dirac particles. The Majorana neutrino magnetic moment is introduced via the effective 
electromagnetic Hamiltonian 
$H_{em}^M=-\frac{1}{4}\nu_L^TC^{-1}\lambda\sigma^{\alpha\beta}\nu_LF_{\alpha\beta}+h.c.$ \cite{Schechter:1981hw}. 
Here $\lambda=\mu-id$ is an antisymmetric arbitrary complex matrix. On the other hand, the corresponding Dirac 
electromagnetic effective Hamiltonian is $H_{em}^D=\frac{1}{2}\bar \nu_R\lambda\sigma^{\alpha\beta}\nu_LF_{\alpha\beta}+h.c.$ 
and in this case, $\lambda=\mu-id$ is an arbitrary complex hermitian matrix \cite{Grimus:2000tq}. 
Experiments are only sensitive to some process-dependent effective neutrino magnetic moment $\mu_\nu$ given 
by a superposition of the matrix elements of  $\lambda$ (see for instance \cite{Canas:2015yoa}). It is this effective moment 
the one that will affect the change of the neutrino helicity $s_{||}$ according to Eq. \eqref{spinequation}, thus we can 
forget about the physics behind the specific value of the effective parameter appearing in this equation and treat on 
equal footing the evolution of $s_{||}$ for both Dirac and Majorana neutrinos.
Experimental limits on this effective magnetic neutrino moment are shown in Table \ref{expdata}.
\begin{figure}
\begin{center}
\includegraphics[width=.45\textwidth]{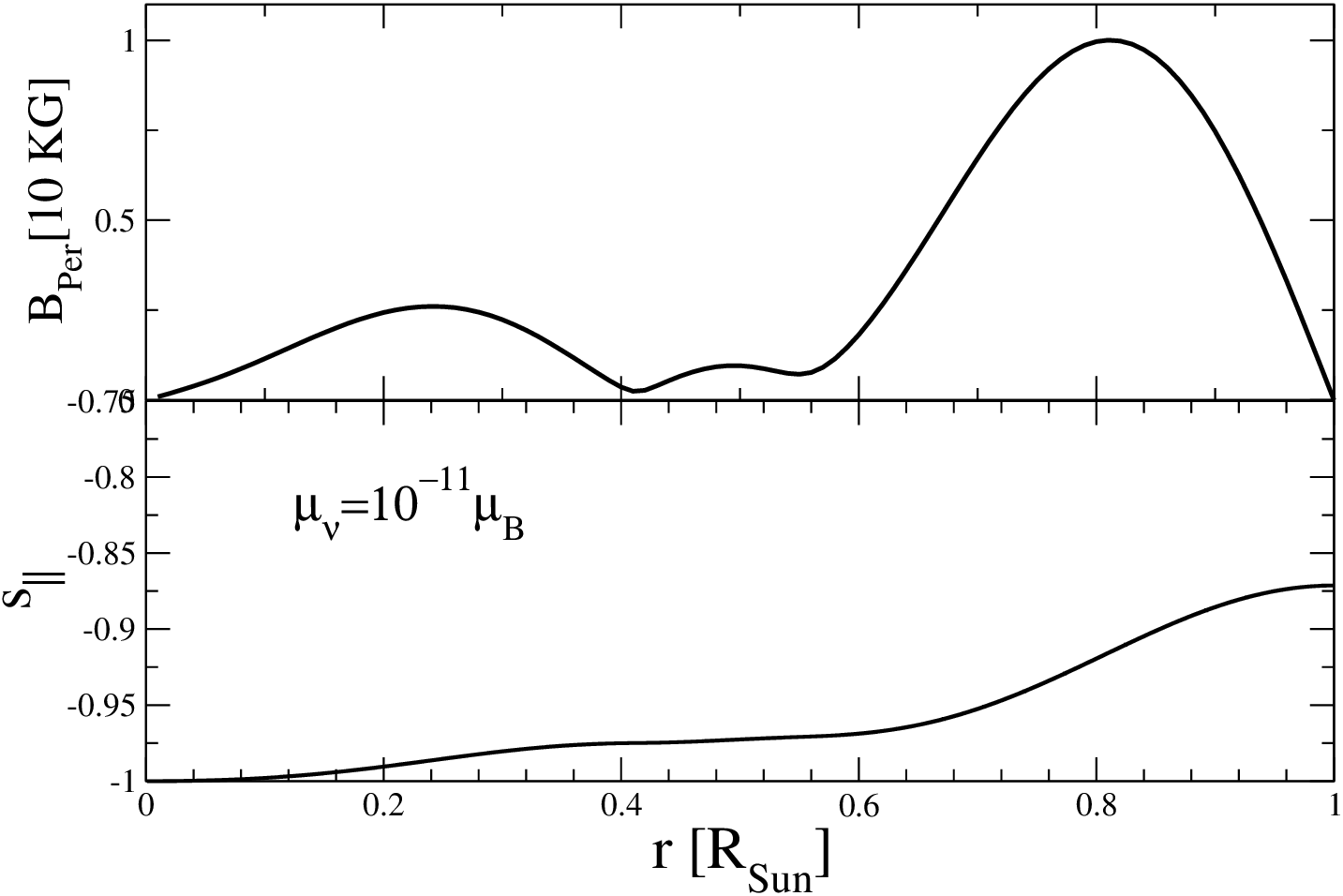}
\caption{Top panel: The magnetic field in the interior of the Sun. Bottom panel: Evolution of the neutrino polarization 
$s_{||}$ as it travels in the solar interior.}\label{Fig2}
\end{center}
\end{figure}

The other relevant ingredient to describe the change of $s_{||}$ is the external magnetic field $B_\perp$ and the next 
sections are devoted to explore two natural sources of neutrinos produced in an environment with magnetic fields: 
the Sun and supernova explosions. 

\begin{center}
\begin{table}\label{expdata}
\caption{Stronger limits for different neutrino effective magnetic moments.}
\begin{tabular}{c|c|c}
\hline\hline
Experiment& Limit & Ref. \\
\hline\hline
GEMMA$_{\textrm{Reactor} \bar \nu_e-e^-}$& $\mu_{\nu_e}< 2.9 \times 10^{-11}\mu_B$ & \cite{Beda:2012zz}\\
LSND$_{\textrm{Accelerator} (\nu_\mu,\bar \nu_\mu)-e^-}$ & $\mu_{\nu_\mu}< 6.8 \times 10^{-10}\mu_B$ & \cite{Auerbach:2001wg}\\
Borexino$_{\textrm{Solar} \bar \nu_e-e^-}$& $\mu_{\nu_e}< 5.4 \times 10^{-11}\mu_B$ & \cite{Arpesella:2008mt}\\
\hline\hline
\end{tabular}
\end{table}
\end{center}

\section{Change of the neutrino helicity in the Sun}\label{sunmagnetic}

The question is whether the Sun magnetic field 
can produce sizeable changes in the helicity of neutrinos to yield a measurable effect in terrestrial detectors or not. 
In order to answer this question, we follow the magnetic profile proposed in \cite{Miranda:2000bi}. It was
obtained by means of full self-consistent and analytical solutions of the magneto-hydrodynamic equations inside the Sun.  
The magnetic field inside the Sun was computed as a family of solutions given by
\begin{equation}
B^k_r(r, \theta)=2\hat B^k\cos \theta\left[1 -
\frac{3}{r^2S_{k}}
\left(\frac{\sin(z_{k}r)}{z_{k}r} - \cos(z_{k}r)\right)\right]~,
\nonumber
\end{equation}
\begin{equation}
B^k_{\theta}(r, \theta)=-\hat B^k\sin\theta  \left[2  +
\frac{3}{r^2S_{k}}\left(\frac{\sin(z_{k}r)(1-(z_{k}r)^2)}{z_{k}r} -
\cos(z_{k}r)\right)\right]~,\nonumber
\end{equation}
\begin{equation}
B^k_{\phi}(r, \theta)=\hat B^k z_{k}\sin \theta\left[r -
\frac{3}{rS_{k}}\left(\frac{\sin(z_{k}r)}{z_{k}r} -
\cos(z_{k}r)\right)\right]~,
\label{soluciones}
\end{equation}
%
where $z_{k}$ denote the roots of the spherical Bessel function, $S_{k}\equiv z_k\sin z_k$ 
and the boundary conditions  $B_{\perp}(r=0)=B_{\perp}(r=R_\odot)=0$ are imposed. The polar angle is $\theta$ and the 
distance $r$ has been normalized to the solar radius $R_\odot$.
The coefficient $\hat B^k(B_{core})$ is given by
\begin{equation}
\hat B^k = \frac{B_{core}}{2(1 - z_{k}/\sin z_{k})}~.
\end{equation}
There is an upper limit on the magnitude of the Solar magnetic moment in the core. It should be
smaller than $30$ G \cite{boruta}. Furthermore, the solar magnetic field in the convective zone 
should be smaller than 100-300 kG \cite{Moreno}. 
Latest analysis suggests a lower value \cite{Miranda:2003yh,Friedland:2005xh,Raffelt:2009mm,Friedland:2002pg} hence we consider a conservative maximum 
value of the magnetic field in the convective zone of $10$ kG.

Taking advantage of the linearity of the magneto-hydrodynamic equations, any linear combination of $B_K$ is also 
a solution, hence  the solar magnetic field is computed as $\vec B=\sum_K c_K \vec B_K$.  More details in 
the method for computing $c_K$ can be found in \cite{Miranda:2000bi}.
Finally, the perpendicular component $B_\perp$, which is relevant to the
neutrino evolution of $s_{||}$, can be computed as $B_{\perp} = \sqrt{B_{\phi}^2 + B_{\theta}^2}$.
In Fig. \ref{Fig2} we show an example of a magnetic field profile for $B_\perp(r)$ in the Sun.

Once we have set the magnetic profile of the Sun, we can solve Eq. \eqref{spinequation} to find $s_{||}(r)$ for different values 
of $\mu_\nu$. As an example, in the bottom panel of Fig. \ref{Fig2} we show $s_{||}$ as a function of the 
radial coordinate $r$ for the upper bound of the neutrino magnetic moment $\mu_\nu=10^{-11}\mu_B$. From this plot, it is clear that the magnetic field in the Sun may produce a considerable change in the neutrino helicity, thus the effects of this change in observables for terrestrial detectors are worthy of study.


\subsection{On the limit on neutrino magnetic moment}
\begin{figure}
\includegraphics[width=0.45\textwidth]{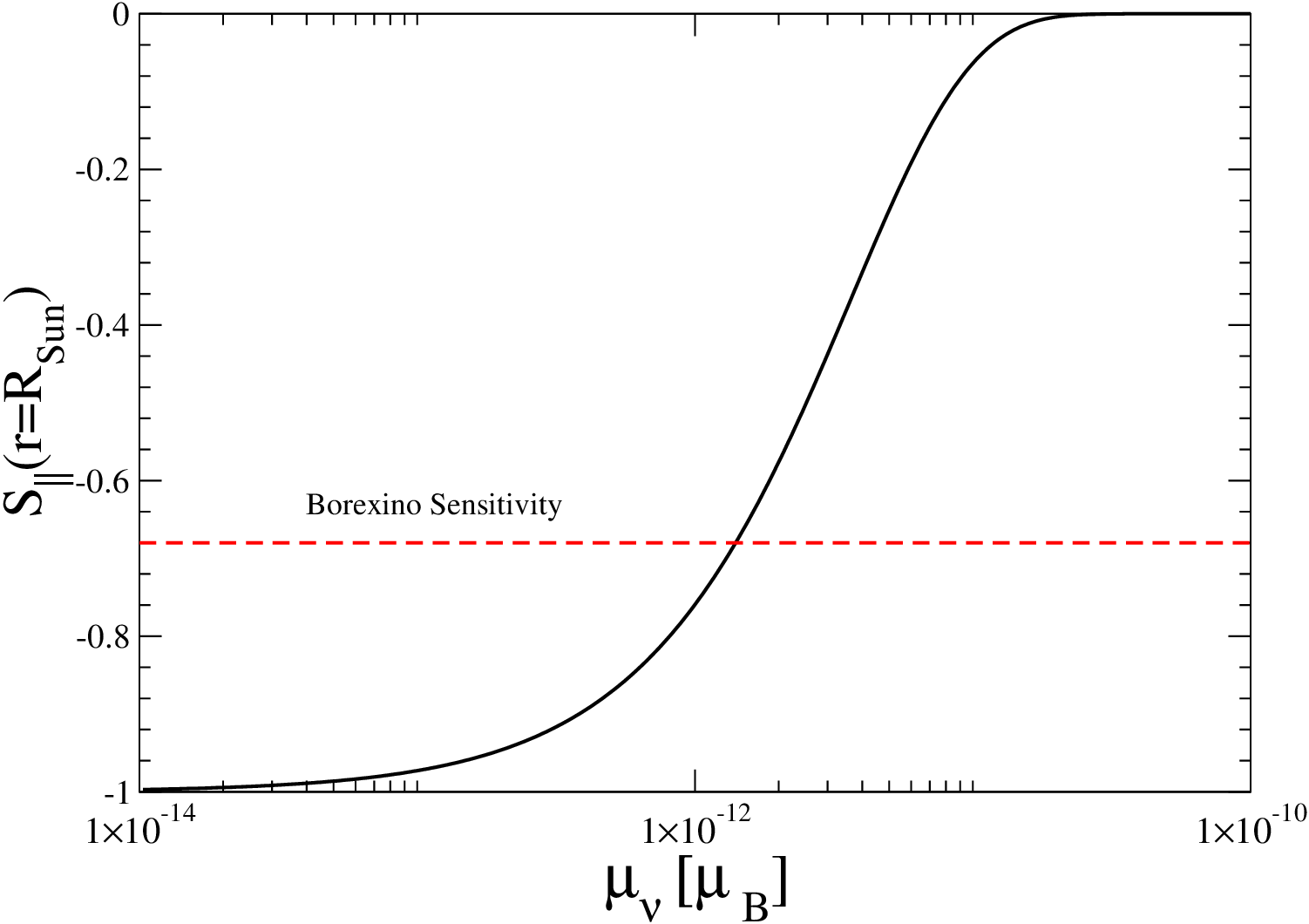}
\caption{The value of $s_{||}$ evaluated at $r=R_\odot$, i.e. the final value of the neutrino helicity when the neutrino 
leaves the Sun for different values of the neutrino magnetic moment $\mu_\nu$.}
\label{Fig4}
\end{figure}
The neutrino magnetic moment $\mu_\nu$ changes the neutrino helicity modifying the expected $\nu$-$e$ elastic 
cross section for solar neutrinos colliding with electrons on Earth. Also, it produces an extra term given by the 
electromagnetic interaction of the neutrino with the electron given by \cite{Bardin:1970wq,Kyuldjiev:1984kz}
\begin{equation}
\frac{d\sigma^{em}}{dT}=\frac{\pi\alpha^2}{m_e^2}\frac{\mu_\nu^2}{\mu_B^2}\left(\frac{1}{T}-\frac{1}{E_\nu}\right)\,.
\label{CrossMag}
\end{equation}
Furthermore, the inclusion of $\mu_\nu$ may change the neutrino probability oscillation. 
Thus,   the number of expected events changes accordingly:
\begin{equation}
N^{M,D}_{theo} = n_t t \phi \int dE_\nu dT P(E_\nu, \mu_\nu)\lambda(E_\nu) 
\frac{d^2\sigma(E_\nu,\mu_\nu,s_{||})^{M,D}}{dE_\nu dT}, \label{events}
\end{equation}
where $n_t$ is the number of targets, $\phi$ the flux, $t$ the observation time and 
$\frac{d^2\sigma(E_\nu,\mu_\nu,s_{||})}{dE_\nu dT}$
includes both s of Eq. \eqref{CrossDirac} or Eq. \eqref{CrossMajorana} and the electromagnetic contribution 
given by Eq. \eqref{CrossMag}.

The $^7Be$ line of the solar neutrino spectrum offers a unique way of testing neutrino-electron cross section. Indeed, 
in this case $E_\nu=0.862$ MeV, the energy distribution  in Eq. \eqref{events}  is a Dirac delta 
function, the probability is only a function of the $\mu_\nu$ and it is easy to  
compute the relative differences in the number of events for Majorana and Dirac neutrinos
\begin{equation}
\frac{|N^M_{theo}-N^D_{theo}|}{N^D_{theo}}=\frac{|\sigma^M(s_\parallel)-\sigma^D(s_\parallel)|}{\sigma^D(s_\parallel)}=D(s_{||}).
\end{equation}
Notice that Eq. \eqref{difference} coincides with the normalized uncertainty in the number of events. 
In this concern, the uncertainty in the data of Borexino number of events translates into an uncertainty in the 
Dirac-Majorana difference $D\left(s_\parallel\right)$, putting an upper bound on this quantity.  

From  the number of neutrino events $N=49\pm1.5\textrm{stat}^{+1.5}_{-1.6}\textrm{syst}$ counts/day/100 ton 
reported by Borexino \cite{Arpesella:2008mt,Bellini:2011rx} we obtain that the Dirac-Majorana difference should 
be less than $4.4\%$. Solving for $D(s_{||})< 0.044$ (see Fig. \ref{differenceBorexino}) we obtain the constraint  
$s_{||} < -0.68$ for the helicity of the neutrinos caught in Borexino detectors.

In order to assess possible effects of the helicity change of solar neutrinos in terrestrial detectors, we need the 
value of the neutrino helicity when it leaves the Sun, $s_{||}(r=R_\odot)$.  This quantity depends on the neutrino magnetic moment.
Assuming that a left handed neutrino was born in the center of the Sun and using the previously described magnetic profile in the Sun, 
we calculate $s_{||}(r=R_\odot)$ as a function of the neutrino magnetic moment.  Our results are shown in Fig. \ref{Fig4}. 
Since most of the magnetic fields between the Earth and the Sun are negligible, neutrinos detected in terrestrial 
experiments will have as an upper bound the polarization given by $s_{||}(r=R_\odot)$. The actual polarization can be lower depending 
on the site in the Sun where the neutrino is produced, but it is expected that most of them are produced in the core of the Sun.
If the terrestrial detector uses as detection channel the $\nu$-$e$ elastic scattering, for a neutrino helicity different from $s_{||}=-1$, there 
will be a difference of neutrino counts due to the different cross sections in Eqs. (\ref{CrossDirac},\ref{CrossMajorana}).

As we mentioned before, the number of Borexino neutrino counts requires $s_{||}$$<$-0.68. The neutrino helicity does 
not change when they are travelling from the Sun to the Earth, hence $s_{||}(r=R_\odot,\mu_\nu)<-0.68$. Solving 
for $\mu_\nu$ (see Fig. \ref{Fig4}), we finally obtain a new upper bound,  $\mu_\nu<1.4 \times 10^{-12}\mu_B$, for 
the neutrino magnetic moment. This is an order of magnitude below the best limits in Table \ref{expdata}, even for the 
conservative values we are using in the modeling of the sun magnetic field.

\section{Supernova prospects}\label{supernova}
Simulations of stellar core collapse and supernova explosions have been making impressive progress, including the 
description of microphysics inputs, i.e. the development of better numerical models  that incorporate the important role 
of nuclear and weak interaction physics \cite{Janka:2006fh,Janka:2012wk}. The new codes include the neutrino 
transport systematically and from those simulations the average $\nu_e$ spectrum emitted by a SN can be
extracted. This spectrum has the quasi-thermal form~\cite{Keil:2002in}:
\begin{equation}
\frac{dN(E)}{dE}=\frac{(1+\alpha)^{1+\alpha}E_{\rm tot}}
{\Gamma(1+\alpha)\bar E^2}
\left(\frac{E}{\bar E}\right)^{\alpha}e^{-(1+\alpha)E/\bar E}\,,\label{SNflux}
\end{equation}
where we use $\bar E=15$~MeV for the average energy, $\alpha=4$ for
the pinching parameter, and $E_{\rm tot}=5\times10^{52}$~erg for the
total amount of energy emitted in $\nu_e$ ~\cite{Keil:2002in}.

\begin{figure}
\includegraphics[width=0.45\textwidth]{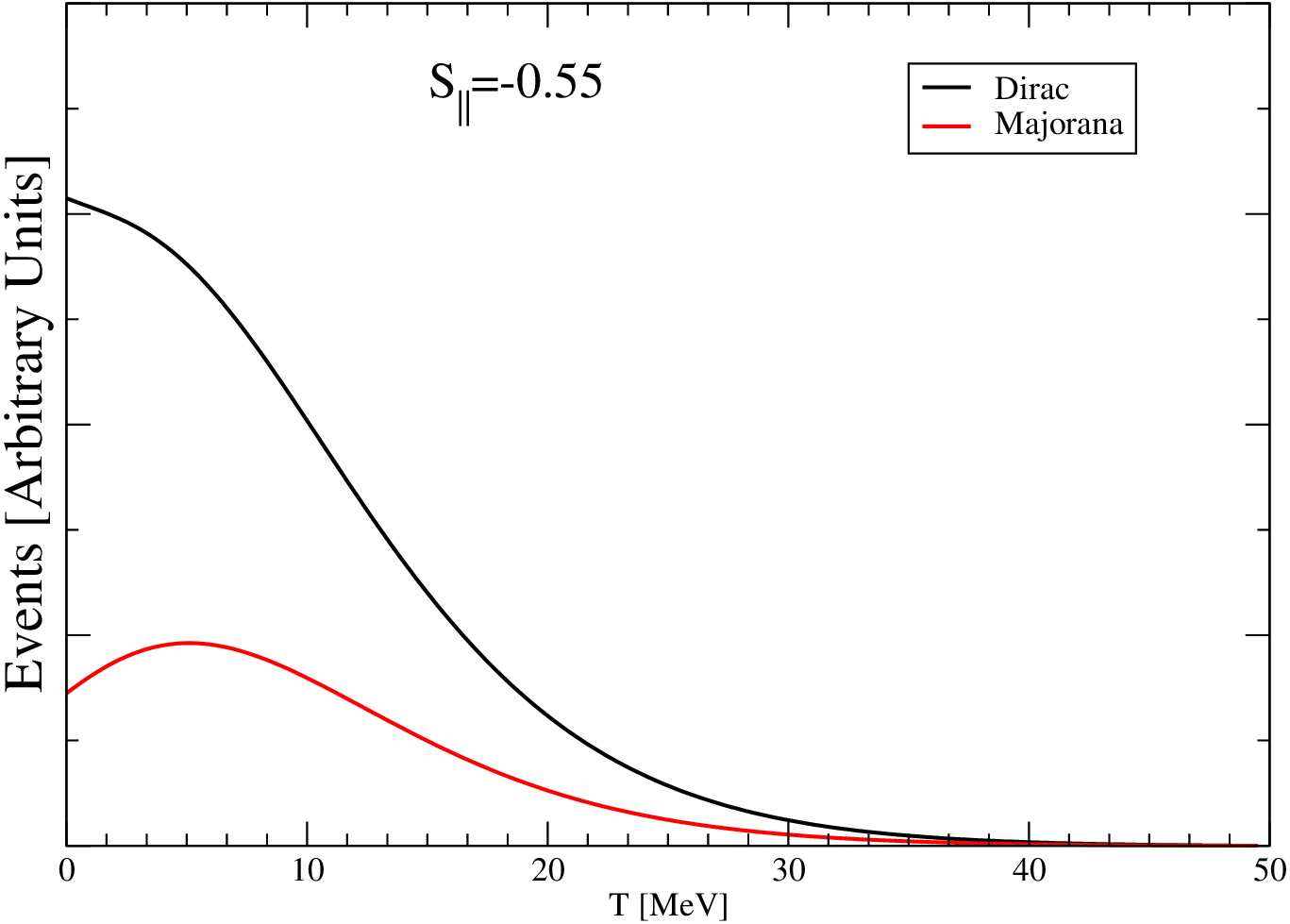}
\caption{Number of supernova events as a function of the recoil energy for Dirac or Majorana neutrinos.}\label{Fig5}
\end{figure}
In addition to the incorporation of neutrino transport, recently, the influence of magnetic fields on stellar core collapse 
and explosion has been explored.
More important, starting with magnetic fields as strong as those predicted by stellar evolution 
($B$$\sim$ $10^9$-$10^{10}$ G), due to turbulent flows, 
the magnetic field may undergo kinematic amplification of some orders of magnitude 
\cite{Obergaulinger2014,Obergaulinger:2011ic}. In some cases,
magnetic field as strong as $10^{15}$ G can emerge for several hundreds of kilometers  \cite{Obergaulinger2014}.
In such violent environment, even a tiny neutrino magnetic moment may induce a change in the neutrino helicity 
$s_{||}$ and thus may lead to changes in the number of events which will be different for Dirac or Majorana neutrinos.
Indeed, a straightforward calculation using Eq. \eqref{spinequation} for a neutrino with original helicity $s_{||}=-1$ 
travelling $100$ km in a constant external magnetic field with strength $B_\perp=10^{15}$ G and $\mu_\nu=10^{-19}\mu_B$, 
which is the magnetic moment predicted for a Dirac neutrino with a mass $m_\nu=1$ eV \cite{Fujikawa:1980yx}, 
yields a final helicity of $s_{||}=-0.55$. We can compute now the number of Majorana or Dirac neutrino events as
\begin{equation}
\frac{dN^{D,M}_{events}}{dT}=A \int dE_\nu \frac{dN(E_\nu)}{dE_\nu} \frac{d^2\sigma^{D,M}(E_\nu,T, S_{||})}{dE_\nu dT},
\end{equation}
where $\frac{dN(E_\nu)}{dE_\nu}$ is the average neutrino spectrum, Eq. \eqref{SNflux}, and the differential neutrino cross 
sections are given by Eq. \eqref{CrossMajorana} and Eq. \eqref{CrossDirac} for Majorana or Dirac neutrinos, respectively. 
The factor $A$ accounts for the number of targets in the detector, the time interval of the detection and the total neutrino flux, 
that depends on the distance from the supernova to the Earth. These parameters  are specific of the supernova explosion 
event thus we are not able to determine the total number of events. Nevertheless, it is possible to perform a qualitative 
analysis of the spectrum in arbitrary units (depending on A) to explore the relative impact of the helicity change due to 
the magnetic fields in a supernova explosion for Majorana and Dirac neutrinos. In  Fig. \ref{Fig5} we plot the distribution 
of the number of events as a function of the electron recoil energy $T$ for $s_{||}=-0.55$ exhibiting clear differences for 
Dirac and Majorana neutrinos even for the small value $\mu_\nu=10^{-19}\mu_B$ considered in the calculation. 


\section{Conclusions}

Massive neutrinos in an environment with magnetic fields can change their helicity $s_{||}$. On the other hand, the 
scattering cross section of  Dirac and Majorana neutrinos on electrons are different if $s_{||}\neq -1$ . In this work we point 
out that nature offer neutrino sources with intrinsic magnetic fields which can provide a sizable change in the neutrino helicity 
and the corresponding difference in the cross section between Dirac and Majorana neutrinos scattering off electrons can be
used to study electromagnetic neutrino properties. 

Considering a particular model of the Solar magnetic field under conservative assumptions, and the number of events for 
solar neutrinos coming from the $^7Be$ line measured by Borexino Collaboration, it is possible to have an improvement 
in the current upper limits of the neutrino magnetic moment of at least one order of magnitude. Furthermore, based on 
accurate numerical modeling of a supernova explosion, we estimate the change in the neutrino helicity of supernova 
neutrinos and show that, even a magnetic moment as small as the predicted by the standard model for Dirac neutrinos, 
 $\mu_\nu=10^{-19}\mu_B$, produces a sizable difference, both in the total counting and in the spectra, of supernova neutrinos 
detected on Earth using $\nu$-$e$ scattering, for Dirac and Majorana neutrinos. 


{\bf Aknowledments:} This work was partially supported by Conacyt projects CB-259228, CB-286651, Conacyt SNI and DAIP.


\begin{thebibliography}{00}

\bibitem{Zralek:1997sa} 
  M.~Zralek,
  Acta Phys.\ Polon.\ B {\bf 28}, 2225 (1997)
  [hep-ph/9711506].
%
\bibitem{Schechter:1981bd}
  J.~Schechter and J.~W.~F.~Valle,
  Phys.\ Rev.\ D {\bf 25} (1982) 2951.
\bibitem{Deshpande:1984sm} 
  N.~G.~Deshpande and G.~Eilam,
  Phys.\ Rev.\ Lett.\  {\bf 53}, 2289 (1984).
  doi:10.1103/PhysRevLett.53.2289
  \bibitem{DellOro:2016tmg} 
  S.~Dell'Oro, S.~Marcocci, M.~Viel and F.~Vissani,
  Adv.\ High Energy Phys.\  {\bf 2016}, 2162659 (2016)
  doi:10.1155/2016/2162659
  [arXiv:1601.07512 [hep-ph]].
  
  \bibitem{Vergados:2016hso} 
  J.~D.~Vergados, H.~Ejiri and F.~?imkovic,
  Int.\ J.\ Mod.\ Phys.\ E {\bf 25}, no. 11, 1630007 (2016)
  doi:10.1142/S0218301316300071
  [arXiv:1612.02924 [hep-ph]].
\bibitem{Singh:2006ad} 
  D.~Singh, N.~Mobed and G.~Papini,
  Phys.\ Rev.\ Lett.\  {\bf 97}, 041101 (2006)
  doi:10.1103/PhysRevLett.97.041101
  [gr-qc/0605153].
\bibitem{Kayser:1981nw}
  B.~Kayser and R.~E.~Shrock,
  Phys.\ Lett.\ B {\bf 112} (1982) 137.
\bibitem{Barranco:2014cda} 
  J.~Barranco, D.~Delepine, V.~Gonzalez-Macias, C.~Lujan-Peschard and M.~Napsuciale,
  Phys.\ Lett.\ B {\bf 739}, 343 (2014)
  doi:10.1016/j.physletb.2014.11.008
  [arXiv:1408.3219 [hep-ph]].


\bibitem{KlapdorKleingrothaus:2000sn}
  H.~V.~Klapdor-Kleingrothaus {\it et al.},
  Eur.\ Phys.\ J.\ A {\bf 12} (2001) 147
  doi:10.1007/s100500170022
  [hep-ph/0103062].
\bibitem{Aalseth:2002rf}
  C.~E.~Aalseth {\it et al.} [IGEX Collaboration],
  Phys.\ Rev.\ D {\bf 65} (2002) 092007
  doi:10.1103/PhysRevD.65.092007
  [hep-ex/0202026].
\bibitem{Auger:2012ar}
  M.~Auger {\it et al.} [EXO-200 Collaboration],
  Phys.\ Rev.\ Lett.\  {\bf 109} (2012) 032505
  doi:10.1103/PhysRevLett.109.032505
  [arXiv:1205.5608 [hep-ex]].
\bibitem{Agostini:2017iyd} 
M.~Agostini {\it et al.},
Nature {\bf 544}, 47 (2017)
doi:10.1038/nature21717
[arXiv:1703.00570 [nucl-ex]].

\bibitem{Gando:2012zm}
  A.~Gando {\it et al.} [KamLAND-Zen Collaboration],
  Phys.\ Rev.\ Lett.\  {\bf 110} (2013) no.6,  062502
  doi:10.1103/PhysRevLett.110.062502
  [arXiv:1211.3863 [hep-ex]].
\bibitem{Arnaboldi:2002du}
  C.~Arnaboldi {\it et al.} [CUORE Collaboration],
  Nucl.\ Instrum.\ Meth.\ A {\bf 518} (2004) 775
  doi:10.1016/j.nima.2003.07.067
  [hep-ex/0212053].
 
  
\bibitem{Barbieri:1991ed}
  R.~Barbieri, G.~Fiorentini, G.~Mezzorani and M.~Moretti,
  Phys.\ Lett.\ B {\bf 259} (1991) 119.
  doi:10.1016/0370-2693(91)90144-F

\bibitem{Semikoz:1996up}
  V.~B.~Semikoz,
  Nucl.\ Phys.\ B {\bf 498} (1997) 39
  [hep-ph/9611383].

\bibitem{Pastor:1997pb}
  S.~Pastor, V.~B.~Semikoz and J.~W.~F.~Valle,
  Phys.\ Lett.\ B {\bf 423} (1998) 118
  [hep-ph/9711316].
\bibitem{Miranda:2003yh}
  O.~G.~Miranda, T.~I.~Rashba, A.~I.~Rez and J.~W.~F.~Valle,
  Phys.\ Rev.\ Lett.\  {\bf 93} (2004) 051304
  [hep-ph/0311014].

\bibitem{Miranda:2000bi}
  O.~G.~Miranda, C.~Pena-Garay, T.~I.~Rashba, V.~B.~Semikoz and J.~W.~F.~Valle,
  Nucl.\ Phys.\ B {\bf 595} (2001) 360
  doi:10.1016/S0550-3213(00)00546-0
  [hep-ph/0005259].


\bibitem{Arpesella:2008mt}
  C.~Arpesella {\it et al.} [Borexino Collaboration],
  Phys.\ Rev.\ Lett.\  {\bf 101} (2008) 091302
  doi:10.1103/PhysRevLett.101.091302
  [arXiv:0805.3843 [astro-ph]].
  
\bibitem{Bellini:2011rx}
  G.~Bellini {\it et al.},
  Phys.\ Rev.\ Lett.\  {\bf 107} (2011) 141302
  doi:10.1103/PhysRevLett.107.141302
  [arXiv:1104.1816 [hep-ex]].

\bibitem{Obergaulinger2014}
M. ~Obergaulinger, H. Th. ~Janka, M.A. Aloy,
Monthly Notices of the Royal Society 445 (2014) 3869.  
[arXiv:1405.7466 [astro-ph]].
  
\bibitem{Obergaulinger:2011ic}
  M.~Obergaulinger and H.~T.~Janka,
  arXiv:1101.1198 [astro-ph.SR].

\bibitem{Dass:1984qc}
  G.~V.~Dass,
  Phys.\ Rev.\ D {\bf 32} (1985) 1239.

\bibitem{Garavaglia:1983wh}
  T.~Garavaglia,
  Phys.\ Rev.\ D {\bf 29} (1984) 387.


\bibitem{Giusarma:2016phn}
  E.~Giusarma, M.~Gerbino, O.~Mena, S.~Vagnozzi, S.~Ho and K.~Freese,
  Phys.\ Rev.\ D {\bf 94} (2016) no.8,  083522
  doi:10.1103/PhysRevD.94.083522
  [arXiv:1605.04320 [astro-ph.CO]].
\bibitem{Ade:2015xua} 
P.~A.~R.~Ade {\it et al.} [Planck Collaboration],
Astron.\ Astrophys.\  {\bf 594}, A13 (2016)
doi:10.1051/0004-6361/201525830
[arXiv:1502.01589 [astro-ph.CO]].

\bibitem{Kraus:2004zw}
  C.~Kraus {\it et al.},
  Eur.\ Phys.\ J.\ C {\bf 40} (2005) 447
  doi:10.1140/epjc/s2005-02139-7
  [hep-ex/0412056].
\bibitem{Lobashev:2001uu}
  V.~M.~Lobashev {\it et al.},
  Nucl.\ Phys.\ Proc.\ Suppl.\  {\bf 91} (2001) 280.
  doi:10.1016/S0920-5632(00)00952-X
\bibitem{Barenboim:1996cu}
  G.~Barenboim, J.~Bernabeu and O.~Vives,
  Phys.\ Rev.\ Lett.\  {\bf 77} (1996) 3299
  [hep-ph/9606218].
\bibitem{Fujikawa:1980yx} 
  K.~Fujikawa and R.~Shrock,
  Phys.\ Rev.\ Lett.\  {\bf 45}, 963 (1980).
  doi:10.1103/PhysRevLett.45.963
  
  
  \bibitem{Bargmann:1959gz} 
  V.~Bargmann, L.~Michel and V.~L.~Telegdi,
  Phys.\ Rev.\ Lett.\  {\bf 2}, 435 (1959).
  doi:10.1103/PhysRevLett.2.435
    
  
\bibitem{Semikoz:1992yw}
  V.~Semikoz,
  Phys.\ Rev.\ D {\bf 48} (1993) 5264
   [Erratum-ibid.\ D {\bf 49} (1994) 6246].
\bibitem{Schechter:1981hw}
  J.~Schechter and J.~W.~F.~Valle,
  Phys.\ Rev.\ D {\bf 24} (1981) 1883
   Erratum: [Phys.\ Rev.\ D {\bf 25} (1982) 283].
  doi:10.1103/PhysRevD.25.283, 10.1103/PhysRevD.24.1883
\bibitem{Grimus:2000tq}
  W.~Grimus and T.~Schwetz,
  Nucl.\ Phys.\ B {\bf 587} (2000) 45
  doi:10.1016/S0550-3213(00)00451-X
  [hep-ph/0006028].
\bibitem{Canas:2015yoa}
  B.~C.~Canas, O.~G.~Miranda, A.~Parada, M.~Tortola and J.~W.~F.~Valle,
  Phys.\ Lett.\ B {\bf 753} (2016) 191
   Addendum: [Phys.\ Lett.\ B {\bf 757} (2016) 568]
  doi:10.1016/j.physletb.2016.03.078, 10.1016/j.physletb.2015.12.011
  [arXiv:1510.01684 [hep-ph]].
  
\bibitem{Beda:2012zz}
  A.~G.~Beda, V.~B.~Brudanin, V.~G.~Egorov, D.~V.~Medvedev, V.~S.~Pogosov, M.~V.~Shirchenko and A.~S.~Starostin,
  Adv.\ High Energy Phys.\  {\bf 2012} (2012) 350150.
  doi:10.1155/2012/350150
\bibitem{Auerbach:2001wg}
  L.~B.~Auerbach {\it et al.} [LSND Collaboration],
  Phys.\ Rev.\ D {\bf 63} (2001) 112001
  doi:10.1103/PhysRevD.63.112001
  [hep-ex/0101039].
\bibitem{boruta} N. Boruta, Astrophys. J., 458 (1996) 832 
\bibitem{Moreno} F. Moreno-Insertis, Astron. \& Astrophys, {\bf 166}
  (1986) 291  
  
  
  
\bibitem{Friedland:2005xh}
  A.~Friedland,
  hep-ph/0505165.
\bibitem{Raffelt:2009mm}
  G.~Raffelt and T.~Rashba,
  Phys.\ Atom.\ Nucl.\  {\bf 73} (2010) 609
  doi:10.1134/S1063778810040058
  [arXiv:0902.4832 [astro-ph.HE]].
\bibitem{Friedland:2002pg}
  A.~Friedland and A.~Gruzinov,
  Astropart.\ Phys.\  {\bf 19} (2003) 575
  doi:10.1016/S0927-6505(02)00255-4
  [hep-ph/0202095].
  
\bibitem{Bardin:1970wq}
  D.~Y.~Bardin, S.~M.~Bilenky and B.~Pontecorvo,
  Phys.\ Lett.\  {\bf 32B} (1970) 121.
  doi:10.1016/0370-2693(70)90602-7
  
\bibitem{Kyuldjiev:1984kz}
  A.~V.~Kyuldjiev,
  Nucl.\ Phys.\ B {\bf 243} (1984) 387.
  doi:10.1016/0550-3213(84)90482-6

\bibitem{Janka:2006fh}
  H.~T.~Janka, K.~Langanke, A.~Marek, G.~Martinez-Pinedo and B.~Mueller,
  Phys.\ Rept.\  {\bf 442} (2007) 38
  doi:10.1016/j.physrep.2007.02.002
  [astro-ph/0612072].
\bibitem{Janka:2012wk}
  H.~T.~Janka,
  Ann.\ Rev.\ Nucl.\ Part.\ Sci.\  {\bf 62} (2012) 407
  doi:10.1146/annurev-nucl-102711-094901
  [arXiv:1206.2503 [astro-ph.SR]].
 
\bibitem{Keil:2002in}
  M.~T.~Keil, G.~G.~Raffelt and H.~T.~Janka,
  Astrophys.\ J.\  {\bf 590} (2003) 971
  doi:10.1086/375130
  [astro-ph/0208035].
  
\end{thebibliography}
\end{document}